\documentclass[10pt,preprint2]{aastex}

\def\sun{\odot}

\begin{document}

\title{The Optical/Near-Infrared Light Curves of SN 2002ap\\
for the First 140 Days after Discovery}

\author{Yuzuru Yoshii\altaffilmark{1,4},
      Hiroyuki Tomita\altaffilmark{2},
   Yukiyasu Kobayashi\altaffilmark{3},
         Jinsong Deng\altaffilmark{2,4},
        Keiichi Maeda\altaffilmark{2},
      Ken'ichi Nomoto\altaffilmark{2,4},
        Paolo A. Mazzali\altaffilmark{6,2,4},
         Hideyuki Umeda \altaffilmark{2},
         Tsutomu Aoki\altaffilmark{1},
           Mamoru Doi\altaffilmark{1},
           Keigo Enya\altaffilmark{1},
       Takeo Minezaki\altaffilmark{1},
    Masahiro Suganuma\altaffilmark{2},
and
    Bruce A. Peterson\altaffilmark{5}
}

\altaffiltext{1}{Institute of Astronomy, School of Science,
                 University of Tokyo, 2-21-1
                 Osawa, Mitaka, Tokyo 181-0015, Japan}
\altaffiltext{2}{Department of Astronomy, School of Science,
                 University of Tokyo, 7-3-1
                 Hongo, Bunkyo-ku, Tokyo 113-0033, Japan}
\altaffiltext{3}{National Astronomical Observatory, 2-21-1
                 Osawa, Mitaka, Tokyo 181-8588, Japan}
\altaffiltext{4}{Research Center for the Early Universe,
                 School of Science, University of Tokyo, 7-3-1
                 Hongo, Bunkyo-ku, Tokyo 113-0033, Japan}
\altaffiltext{5}{Mount Stromlo Observatory,
                 Research School of Astronomy and Astrophysics,
                 The Australian National University,
                 Weston Creek P. O., A. C. T. 2611, Australia}
\altaffiltext{6}{Osservatorio Astronomico, Via Tiepolo 11,
                 34131 Trieste, Italy}


\begin{abstract}

Supernova (SN) 2002ap in M74 was observed in the $UBVRIJHK$ bands
for the first 40 days following its discovery (2002 January 29)
until it disappeared because of solar conjunction, and then in
June after it reappeared. The magnitudes and dates of peak
brightness in each band were determined. While the rate of
increase of the brightness before the peak is almost independent
of wavelength, the subsequent rate of decrease becomes smaller
with wavelength from the $U$ to the $R$ band, and is constant at
wavelengths beyond $I$.  The photometric evolution is faster than
in the well-known ``hypernovae'' SNe~1998bw and 1997ef, indicating
that SN~2002ap ejected less mass. The bolometric light curve of
SN~2002ap for the full period of observations was constructed. The
absolute magnitude is found to be much fainter than that of
SN~1998bw, but is similar to that of SN~1997ef, which lies at the
faint end of the hypernova population.  The bolometric light curve
at the early epochs was best reproduced with the explosion of a
C+O star that ejects $2.5~M_\sun$ with kinetic energy $E_{\rm
K}=4\times 10^{51}~{\rm ergs}$. A comparison of the predicted
brightness of SN~2002ap with that observed after solar conjunction
may imply that $\gamma$-ray deposition at the later epochs was
more efficient than in the model. This may be due to an asymmetric
explosion.

\end{abstract}
\keywords{supernovae: general---supernovae: individual (SN 2002ap)---supernovae: photometry}


\section{Introduction}

The discovery of hypernovae, i.e., supernovae (SNe) with explosion
energies exceeding that of normal SNe, and possibly linked with
gamma-ray bursts, has motivated the creation of a close network
between transient-object hunters, follow-up observers, and model
builders. This network successfully functioned for the recent
energetic Type~Ic SN~2002ap, which appeared in the outer part of
the nearby spiral galaxy M74 ($m-M=29.5_{-0.2}^{+0.1}$, $D\approx
7.9$~Mpc; Sharina, Karachentsev \& Tikhonov 1996; Sohn \& Davidge
1996). After its discovery on January 29 by Y. Hirose and
confirmation the following day (Nakano et~al. 2002), intensive
world-wide follow-up observational programs (e.g., Gal-Yam, Ofek
\& Shemmer 2002) were started, accumulating photometric and
spectroscopic data. Theoretical models (e.g., Mazzali et~al. 2002)
have been used to constrain the physical properties of SN~2002ap.

Having received notification of the rare opportunity of a nearby
SN event soon after discovery, multi-wavelength monitoring of the
brightness of SN~2002ap in the $UBVRIJHK$ bands was started with
the multi-color imaging photometer (MIP) mounted on the MAGNUM
2m telescope (Yoshii et~al. 2002), which is located at a
height of 3055m at the summit of Haleakala, on the Island of Maui,
Hawaii. The MAGNUM (\underline{M}ulti-color \underline{A}ctive
\underline{G}alactic \underline{Nu}clei \underline{M}onitoring)
Project was designed to monitor many AGNs at wavelengths from
the UV/optical to the near infrared (NIR) and to determine a
redshift-independent distance to the AGNs from which the
cosmological parameters could finally be determined (Yoshii 2002).

In this paper, the full set of $UBVRIJHK$ imaging observations
of SN~2002ap taken during the first 40 days following its discovery
until it disappeared because of solar conjunction,
and then in June after it reappeared, is presented, including the first NIR
light curves
\footnote{A part of the data presented here was previously used
in Mazzali et~al. (2002) for the sake of theoretical modelling.}.
These light curves are analyzed and compared to those of the
hypernovae SNe 1998bw and 1997ef.
The bolometric light curve, constructed for
dates up to March 11, and then combined with one point in June,
is compared with that synthesized using theoretical models.

\section{Imaging Photometry}

\subsection{Observations}

The MAGNUM telescope has a 2m primary mirror and Rithey-Chr\'etien
optics giving a field view of 33 arcminutes (Kobayashi et~al. 1998a).
The MIP camera, installed at the focal point of the bent-Cassegrain
focus, contains an SITe CCD ($1024\times 1024$ pixels, 0.277 arcsec/pixel)
and an SBRC InSb array ($256\times 256$ pixels, 0.346 arcsec/pixel).
A dichroic beam splitter enables simultaneous imaging through optical
($UBVRI$) and NIR ($ZJHK'KL'$) filters (Kobayashi et~al. 1998b).
The effective area of each frame over which the light was received
by the detector is $119\times 119$ arcsecs for the CCD and
$88.5\times 88.5$ arcsecs for the InSb.  Observations of SN~2002ap
are on the Johnson-Cousins system after correction for color-terms
and color-offsets.

SN~2002ap ($\alpha_{2000}= 01^h 36^m23.^s85,
\delta_{2000}=+15^{\circ}45'13.''2$, Nakano et al. 2002) was
observed in the $UBVRIJHK$ bands starting on February 2 and until
March 11, 2002 before solar conjunction. Observations performed
during June 2002, the first month after solar conjunction, are
also reported. Given the inferred explosion date, January 28
(Mazzali et~al. 2002), our observation began only 5 days after the
explosion. A nearby star ($\alpha_{2000}= 01^h 36^m19.^s49,
\delta_{2000}=+15^{\circ}45'20.''8$), known to be a double star,
was of comparable brightness to SN~2002ap in the same CCD frame,
so that it was observed as a reference star for $UBVRI$.

Although SN~2002ap was only visible at low elevations ($40-20$
degrees), either after sunset (for the observations before solar
conjunction) or before sunrise (for the observations after solar
conjunction), MIP's efficient splitting of the incident light
allowed the successful completion of simultaneous multi-color
imaging within an hour or so.  Because of such simultaneous
imaging, telescope dithering required primarily for $JHK$ was also
performed for $UBVRI$ using 3 10-arcsec steps. A typical exposure
time for one step was 120 sec for $U$, 40 sec for each of $BVRI$,
and 10 sec for each of $JHK$.  However, for the observations in
June, the exposure time was prolonged within the saturation limits
of the reference star. Typical seeing sizes of the stellar images
were 1.0-1.7 arcsec in $V$ and 0.85-1.3 arcsec in $K$,
respectively.

Our observational strategy was to observe SN~2002ap, the infrared
standard stars for $JHK$ (Hunt et~al. 1997) and the optical
standard stars for $UBVRI$ (Landolt 1992), irrespective of whether
the night was photometric or not. Those observations were performed
by interrupting the scheduled queue of our original program of
observing AGNs. At the end of each night, dome-flat images for
$BVRIJHK$ and infrared dark images were obtained.  However,
twilight-flat images for $U$ were obtained on other days.

\subsection{Image reduction}

Image reduction was performed using our package of IRAF-based
automated reduction software (MAGRED), which was developed
for the MAGNUM Project.  A simple pattern of noises, which was
overlayed on the images by interference from the radio broadcast
station near the telescope site, was removed from the bias
region of each CCD frame. Then, for the CCD data reduction,
nonlinear corrections, bias subtraction, flat fielding based
on the dome flat (twilight flat for $U$), and interpolation
among bad pixels were applied to each frame in a standard manner.
Sky subtraction was not employed because the difference between
sky flat and dome flat was negligibly small except near the edge
of each frame.

For the InSb data reduction, the noise pattern was removed from
the image region of each frame, because the bias region is absent
in the InSb. Then, similar corrections as above were made, but sky
subtraction was done by using the sky flat taken at a time close to
the observation of SN~2002ap.   Flat images for $H$ and $K$ were
generated as the difference between the dome-flat images taken with
the light on and off, after correcting for illumination by comparing
with the sky flat.  Flat images for $J$ were taken similarly except
that dark images were used instead of dome-flat images taken with the
light off.

Multi-aperture photometry was used throughout, since SN~2002ap is
located far from the center of M74, in a region where the sky
background is sufficiently flat.  An aperture size of 11 arcsec
was chosen as a compromise between smaller apertures which
reliably increase the S/N ratio, and larger apertures which ensure
constant and stable measured magnitudes of the reference double
star. However, in order to enhance the S/N ratio, a smaller
aperture was adopted for the $U$-band images taken on March $4-5$
(8.3 arcsec), and for the images taken in June (5.5 arcsec for
$BVRI$ and 8.3 arcsec for $JHK$).

Three CCD frames of SN~2002ap were obtained each night while
dithering the telescope position, and aperture photometry
was performed for each frame. The adopted aperture magnitude
of SN~2002ap is the median of the three.  However, for the $U$-band frames
taken after February 21, the three frames were stacked before aperture
photometry was performed.
Figure 1 shows a series of color-coded images taken on eight
different nights (February 2, 4, 8, 14, 16, 18, 22, and March 5),
which give a visual impression of the optical color evolution
of SN~2002ap.

Since each CCD frame is wide enough to cover both SN~2002ap
and the reference star, the aperture magnitude of SN~2002ap was
obtained via differential photometry between SN~2002ap and
the reference star in the same CCD frame (Enya et~al. 2002).

The magnitude calibration of the reference star in the CCD frame
was carried out only on photometric nights, i.e. when the
dispersion in instrumental magnitudes of SN2002ap and that for the
reference star were comparable to the S/N ratio, and also when it
was judged to be a fine day from the MAGNUM cloud monitor data
(Suganuma et~al. in preparation) and by eye inspection of the
weather.  Whenever optical standard stars were observed more than
twice at different elevations on a photometric night, the airmass
dependence and the magnitude zero points in each band were
obtained, so that the aperture magnitudes of the reference star
were calibrated. Their respective medians and number of
calibrations are $U(4)=14.13\pm$0.04, $B(5)=13.83\pm$0.02,
$V(5)=13.07\pm$0.01, $R(5)=12.61\pm$0.015, and
$I(3)=12.15\pm$0.01, where the error is estimated as half of the
difference between the brightest and faintest magnitudes for
multiple calibrations in each band.

On the other hand, the InSb frame of the $JHK$ observations is
not wide enough to include any reference star of brightness
comparable to SN~2002ap. Therefore, the aperture $JHK$ magnitudes
of SN~2002ap in individual image frames were calibrated using
standards-based photometry, and for each band the median was
adopted as the aperture magnitude of SN2002ap.

\subsection{Observed light curves}

Figure 2 shows the $UBVRIJHK$ light curves based on the data taken
on 14 nights during the period February 2 to March 11, 2002,
until SN~2002ap disappeared behind the sun.
These data are listed in Table 1, together with the data for 5
nights in June after solar conjunction. For the purpose of
comparison, data reported by other authors are also shown in Figure 2.

The magnitude error in each optical band is estimated as
$\sigma_e^2=\sigma_{sn}^2+\sigma_{ref}^2+\sigma_*^2$, where
$\sigma_{sn}$ and $\sigma_{ref}$ are the photometric errors of
SN~2002ap and the reference star, respectively, which were
estimated from the output of IRAF PHOT, and $\sigma_*$ is the
magnitude calibration error of the reference star given in the
previous section.  For June data taken when SN~2002ap faded, the
use of a small aperture for both SN~2002ap and the reference
double star introduced some small error because of their different
brightness profiles.  This error was estimated to be 0.01 mag and
was incorporated in the quadrature.

The magnitude error in each NIR band is estimated as
$\sigma_e^2=\sigma_{sn}^2+\sigma_{std}^2+\sigma_{zp}^2$, where
$\sigma_{sn}$ and $\sigma_{std}$ are the photometric errors of
SN~2002ap and the standard stars, respectively, and $\sigma_{zp}$
is the zero-point uncertainty of the standard stars caused by
time-variant atmospheric conditions. The error $\sigma_{sn}$ is
basically given by the IRAF magnitude dispersion of SN~2002ap
divided by the square root of the number of frames obtained during
the same night.  However, since the error given this way is
smaller than the standard deviation of the instrumental magnitudes
of SN~2002ap, additional errors likely arising from the flat
fielding and the reduction procedure must also be added in
quadrature. We adopted $\sigma_{std}=0.01$ mag because all
standard stars were observed with enough S/N.  We adopted
$\sigma_{zp}=0.02$ mag for clear nights and 0.05 mag for poor
conditions. The errors are larger when thin patchy clouds were
present.

The lines in Figure 2 are fits to the data using cubic splines
weighted by observational errors.  Larger weights were given to
the $UBVRI$ fitting, while relatively small weights were given
to the $JKH$ fitting.  In this fitting procedure, we combined
our own data (filled circles) with other $B$ and $H$ data (open
circles) reported in the IAUCs and Gal-Yam et al.'s $I$ data
(crosses) from the Wise Observatory.  Except for the $I$ band,
where we have no data before February 14, we could obtain light
curves covering the entire period using only our own data.

The date and magnitude of the epoch of peak brightness were
determined from the spline curve for each of $UBVRIJHK$.
The uncertainty of the epoch of maximum was defined as the
range of epochs when the magnitude is fainter than the peak
magnitude by no more than 0.01 mag. The results are shown in
Table 2.  These peak dates agree well with those derived
previously by Gal-Yam et~al. (2002).

The wavelength-dependent slope of the declining part of the
light curve, $dm_\lambda/dt$ (mag/day), was also determined.
These results are also given in Table 2.

\section{Bolometric Light Curve}

Bolometric light curves of SNe are very valuable for constructing
theoretical models. Our photometric data, covering from $U$ to
$K$, enable us to build the bolometric light curve for SN~2002ap.
For days when data in one or two bands are not available from
MAGNUM measurements, we interpolated the data taken on adjacent
dates from MAGNUM and other sources (Gal-Yam et~al. 2002, IAUCs
and VSNET
\footnote{http://vsnet.kusastro.kyoto-u.ac.jp/vsnet/SNe/sn2002ap.html}).

First, all $UBVRIJHK$ magnitudes were converted into absolute
monochromatic fluxes.  Various sets of conversion factors for the
standard system are available in the literature.  As differences
between these various sets are marginal, we adopted the values
recently compiled by Bessell (2001). We adopted a combined
Galaxy-M74 reddening of $E(B-V)=0.09\pm 0.01$, as determined
from the \ion{Na}{1}~D equivalent widths in a high resolution
spectrum (Takada-Hidai, Aoki \& Zhao 2002), and a distance modulus
$m-M=29.5_{-0.2}^{+0.1}$, combining the estimates of Sharina et~al.
(1996) and Sohn \& Davidge (1996) from the brightest supergiants.

Monochromatic fluxes were then integrated over frequency with a
cubic spline interpolation technique. Those on February 8 are
plotted in Figure 3 as an example. We assumed zero flux at both
integral limits, i.e. at the blue edge of the $U$ band and at the
red edge of the $K$ band. The errors introduced by this assumption
are much less than 10\%. An early UV flux between 245 - 320 nm was
measured with XMM-Newton on February 3 (Soria \& Kong 2002).
Extending the integration to this band, we found it contributes
only 4\% of the total flux. The NIR flux, on the other hand,
accounts for a significant fraction of the total flux, ranging
from 20\% on February 2 to 40\% on March 10, and being around 20\%
in June. This highlights the importance of $JHK$ photometry.

Uncertainty in the bolometric light curve also comes from the
non-stellar nature of SN spectra, which are dominated by very
broad lines.  The true bolometric flux can only be obtained
from spectrophotometry.  We compared our results with the model
UV-optical-NIR spectrophotometry derived from the best-fit
synthetic Monte Carlo spectra of Mazzali et~al. (2002), and found
that the differences between the two datasets are less than
0.2 mag.  The total uncertainty in our bolometric light curve,
including the uncertainty in the distance modulus, is therefore
estimated to be $\lesssim 0.3$ mag.

We list in Table 3 our bolometric magnitudes, $M_{\rm bol}$,
along with the bolometric corrections, ${\rm B.C.}\equiv M_{\rm
bol}-M_{\rm V}$. They are shown in Figure 4 as filled circles.
We adopted an explosion date of January 28 (Mazzali et~al. 2002).
We also plot the spectrophotometric point on January 30 as
an open circle, and four unfiltered CCD magnitude points taken
within 1 day of the discovery (Nakano et~al. 2002; VSNET) as crosses.
These roughly define the very early phase of the bolometric light curve.

\section{Discussion}

\subsection{Comparison with other hypernovae}

The multi-band light curves of SN~2002ap are narrower than those
of the well-studied Type~Ic hypernovae, SNe~1998bw and 1997ef.
This is shown in Figure 4, where the bolometric light curve of
SN~1998bw (Patat et~al. 2001), shifted down by 1.85~mag for the
sake of comparison, and that of SN~1997ef (Mazzali et~al. 2000),
are plotted as open circles and filled stars, respectively.
The light curve of SN~2002ap declines more rapidly after peak,
indicating that the trapping of $\gamma$-rays released in the
$^{56}$Ni$\rightarrow$$^{56}$Co$\rightarrow$$^{56}$Fe decays is
less efficient in SN~2002ap, and that the radiation generated by
trapped $\gamma$-rays can escape more easily.
This suggests that the ejecta of SN~2002ap are less massive.

Figure 4 also shows that SN~2002ap was about 1.8~mag fainter than
SN~1998bw at peak, and about as bright as SN~1997ef. The
brightness of the bolometric peak, $\sim -16.9$ mag, and its
epoch, $\sim$ day 11, are similar to those of SN~1994I (Nomoto
et~al. 1994). This indicates that a similar amount of $^{56}$Ni,
$\sim 0.07~M_\sun$, was produced by the two SNe, as suggested by
theoretical models (Mazzali et~al. 2002). SN~2002ap is not as
powerful an explosion as SN~1998bw, which ejected $\sim
0.4-0.7~M_\sun$ of $^{56}$Ni (Nakamura et~al. 2001), but rather
more similar to SN~1997ef, which ejected $\sim 0.1~M_\sun$.

Figure 5 shows a comparison of the colors of SN~2002ap with those
of SN~1997ef. In the figure, the time axes of SNe~1998bw and
1997ef are compressed by a factor of 1.5, which clearly shows that
SN~2002ap evolves about 1.5 times faster than SNe~1998bw and
1997ef in terms of colors. This is in agreement with the finding
by Kinugasa et~al. (2002) that the evolution of the spectra of
SN~2002ap proceeds at about 1.5 times the rate of SN~1997ef, which
again suggests that SN~2002ap ejected less material than
SNe~1998bw and 1997ef.

\subsection{NIR light curves}

Our NIR photometry is the most intensive for any Type Ic SN.

In Figure 6 we compare the absolute $JHK(K')$ light curves of
SN~2002ap with those of the other four Type Ic SNe with published
NIR photometry. These are SNe~1983I (Elias et~al. 1985), 1983V
(Clocchiatti et~al. 1997), 1994I (Rudy et~al. 1994; Grossan et~al.
1999) and 1998bw (Patat et~al. 2001). Among these, the hypernova
SN~1998bw stands out for being about 1.5 magnitudes brighter. The
$J$-band light curve of SN 1983V is significantly broader than
that of SN~2002ap, consistent with its designation as a "slow"
Type Ic in the optical (Clocchiatti et~al. 1997). For the
otherwise well-observed SN~1994I, there is only one data point in
each of $J$ and $H$, and two in $K'$. These show that the NIR
light curves of SN~1994I were probably much narrower than those of
SN~2002ap. The data of SN~1983I seem to connect smoothly with
those of SN~2002ap before day 50 and after day 130, but this may
be the result of our arbitrarily assigning the epochs.

Better understanding of the $JHK$ photometry of Type Ic SNe
and hypernovae requires theoretical modelling in the NIR,
which is beyond the scope of this paper.

\subsection{Comments on theoretical models}

Based on spectrum synthesis, Mazzali et~al. (2002) have proposed
as a viable model for SN~2002ap the energetic spherical explosion
of a C+O star, ejecting $2.5-5~{M_\sun}$ of material with a
kinetic energy of $4-10\times 10^{51}~{\rm ergs}$. Their favorite
model is one with $M_{\rm ej}=2.5~{\rm M_\sun}$ and $E_{\rm
K}=4\times 10^{51}~{\rm ergs}$. This is the explosion of a
$5~{M_\sun}$ C+O star, leaving a compact remnant of mass $M_{\rm
rem}=2.5~{\rm M_\sun}$, possibly a black hole. Such a C+O
star is formed in a He core of mass $M_\alpha=7.0~M_\sun$,
corresponding to a main-sequence mass $M_{\rm ms}\approx
20-25~M_\sun$. The light curve obtained from this model, which is
shown in Figure 4 as a solid line, fits the bolometric light curve
very well before day 50. \footnote{The model light curve in the
first day shown in Figure 4 of Mazzali et~al. (2002) is in
disagreement with the early unfiltered photometry.
That was caused by a numerical error and has been corrected.}

However, the model light curve deviates from the observations at
later phases, lying 0.7 mag above the bolometric point in June.
This suggests that the $\gamma$-ray optical depth at later phases
may be underestimated by the Mazzali et~al. (2002) model. A similar
problem occurred in the case of SN~1998bw, where the models that best
reproduce the observed light curve near the peak underestimate its
brightness around day 150 (Nakamura et~al. 2001) by $\sim 1$ mag.
Maeda et~al. (2003) recently proposed a general parameterized two-component
ejecta model and successfully reproduced the light curves of SNe 1998bw
and 2002ap at both peak and later phases. They suggested that the denser
inner component introduced to increase the $\gamma$-ray optical depth
at later phases is the natural outcome of jet-driven hypernovae explosions
(Maeda et~al. 2002).

H\"oflich, Wheeler \& Wang (1999) introduced aspherical but
low-energy and low-mass ejecta to reproduce the light curve of
SN~1998bw. In fact, it is not difficult to find low-energy
solutions, aspherical or spherical, for hypernova light curves by
decreasing the mass of the ejecta, (e.g. Iwamoto et al. 1998).
However, we must emphasize that it is the unusually broad spectral
features that make hypernovae distinct from normal SNe, and that
these strongly require high kinetic energies in both spherical
models (Iwamoto et~al 1998) and aspherical ones (Maeda et~al.
2002). It is unclear whether the low-energy model of H\"oflich
et~al. (1999) can reproduce the broad lines in SN~1998bw. Wang
et~al. (2002) also warned that very broad line features might have
been observed in the spectra of other Type Ib/Ic SNe, so far
classified as normal, if they had been observed at a time as early
as SN 2002ap.  Obviously, future early observations are necessary
to verify this speculation.

As for SN~2002ap, Wang et al. (2002) interpreted the weak
polarization in SN~2002ap as due to an asymmetry ($\sim 10$\%) in
the ejecta. Spectropolarimetry was also obtained by Kawabata et
al. (2002) and Leonard et al. (2002) at somewhat later epochs.
Their estimate of the degree of asymmetry is similar, and they
discovered an intriguingly fast evolution of the polarization
angle. Our estimate of the kinetic energy of SN~2002ap may change
if asymmetry is introduced. In the case of SN~1998bw,
concentrating the kinetic energy along the line of sight led Maeda
et al. (2002) to reduce its value to $10\times 10^{51}~{\rm
ergs}$. However, Wang et al. (2002) suggested that SN~2002ap is
likely viewed near the equatorial plane, unlike SN~1998bw. In the
aspherical models of Maeda et al. (2002), this would actually lead
to a larger, not a smaller kinetic energy. In any case, we do not
expect that the degree of asphericity, about 10\%-20\%, will
significantly affect the global values derived in spherical
symmetry.

Berger, Kulkarni \& Chevalier (2002) claimed that the Mazzali
et~al. (2002) model places some $10^{48}$~ergs kinetic energy at
$v > 10^5$~km~s$^{-1}$, which is too large, considering the weak
radio emission observed. We note, however, that Mazzali et~al.
(2002) constructed their model by fitting the optical observations
and hence the density structure they used only extended up to
60,000km s$^{-1}$. It is therefore not justified to extrapolate it
to above $v > 10^5$~km~s$^{-1}$, where the material has too low a
density to make an observable contribution to the optical spectra
and light curves, except for polarimetry (Kawabata et~al. 2002).

\acknowledgements

This work has been supported partly by the Grand-in-Aid of
Scientific Research (10041110, 10304014, 12640233, 14047206, 14253001,
14540223) and COE Research (07CE2002) of the Ministry of Education,
Science, Culture, and Sports of Japan. J.D. is supported by a Japan
Society for the Promotion of Science Postdoctoral Fellowship for
Foreign Researchers.  Y.Y. is very grateful to Lena Okajima for
preparing the manuscript.
\newpage



\clearpage

\figcaption[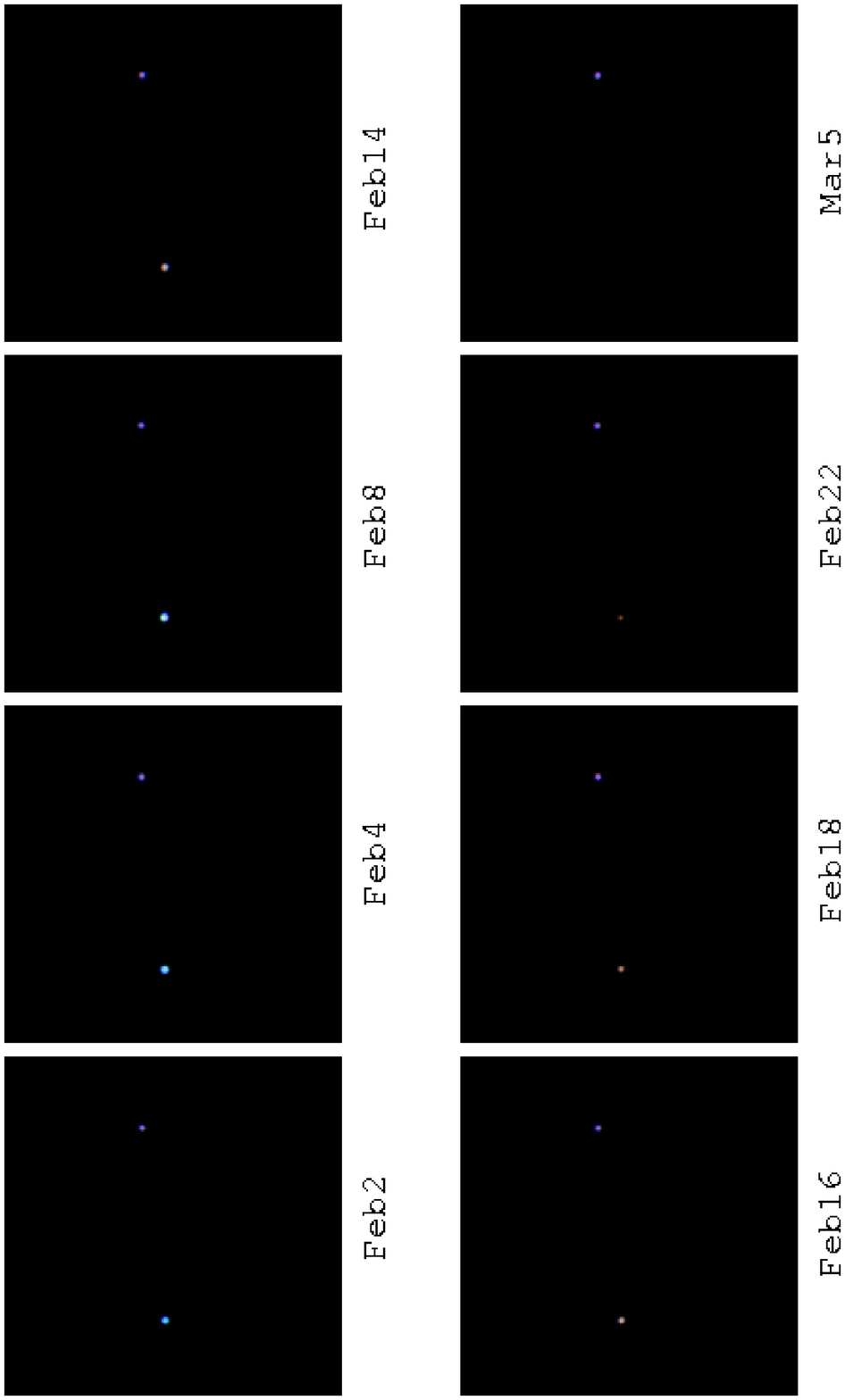]{ A series of optical color-coded
images taken on different nights before and after the epoch
of peak brightness of SN~2002ap, which is around February 8
for the $V$ band.  Each image with view size of 105 arcsec
$\times$ 105 arcsec is made from the images of three
$B$, $V$, and $R$ filters.  While the reference star with red
color on the right stays constant in color and brightness,
SN~2002ap with blue color on the left becomes brighter until
February 8, thereafter becoming redder and fainter.
}

\figcaption[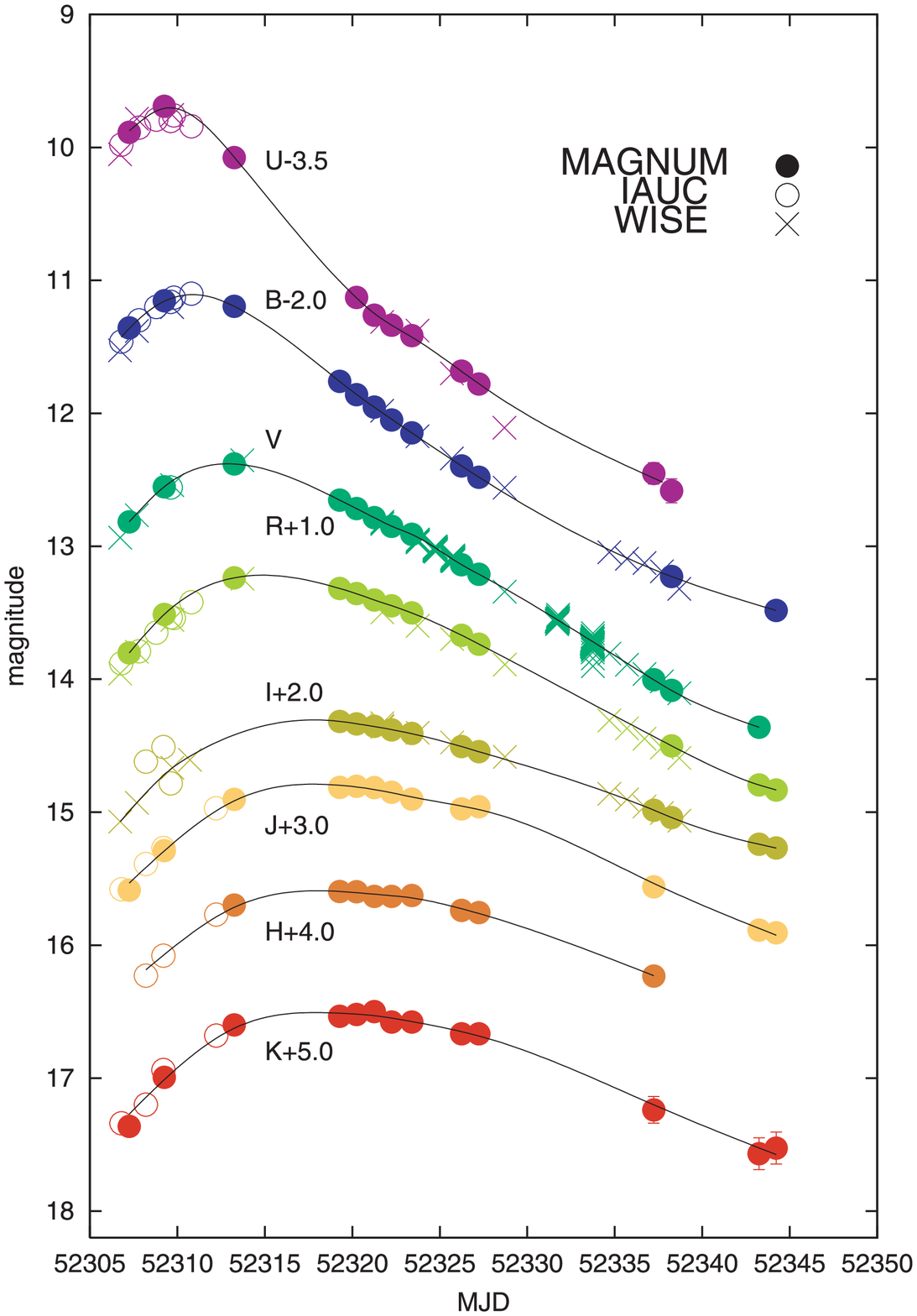]{ Multi-band light curves of SN~2002ap in the
$UBVRIJHK$ bands for the observations before solar conjunction.
The magnitude scales are shifted vertically in order to avoid
overlap. Filled circles show our data, and other symbols show the
data reported by other authors (open circles by Mattila, Meikle,
\& Chambers 2002, Hasubick \& Hornoch 2002, Motohara et~al. 2002,
Riffeser, Goessl, \& Ries 2002, and Szokoly 2002; crosses by
Gal-Yam et al. 2002). The error in each band corresponds to the
combined uncertainty in calibrating the magnitude of SN~2002ap and
the reference star ($UBVRI$) or standard stars ($JHK$).  The
errors for March $4-11$ are larger, as thin patchy clouds were
present. Solid lines are cubic splines fitted mostly to our data
weighted by observational errors.  See the text for details of the
fitting procedure. \label{fig_lightcurve} }

\figcaption[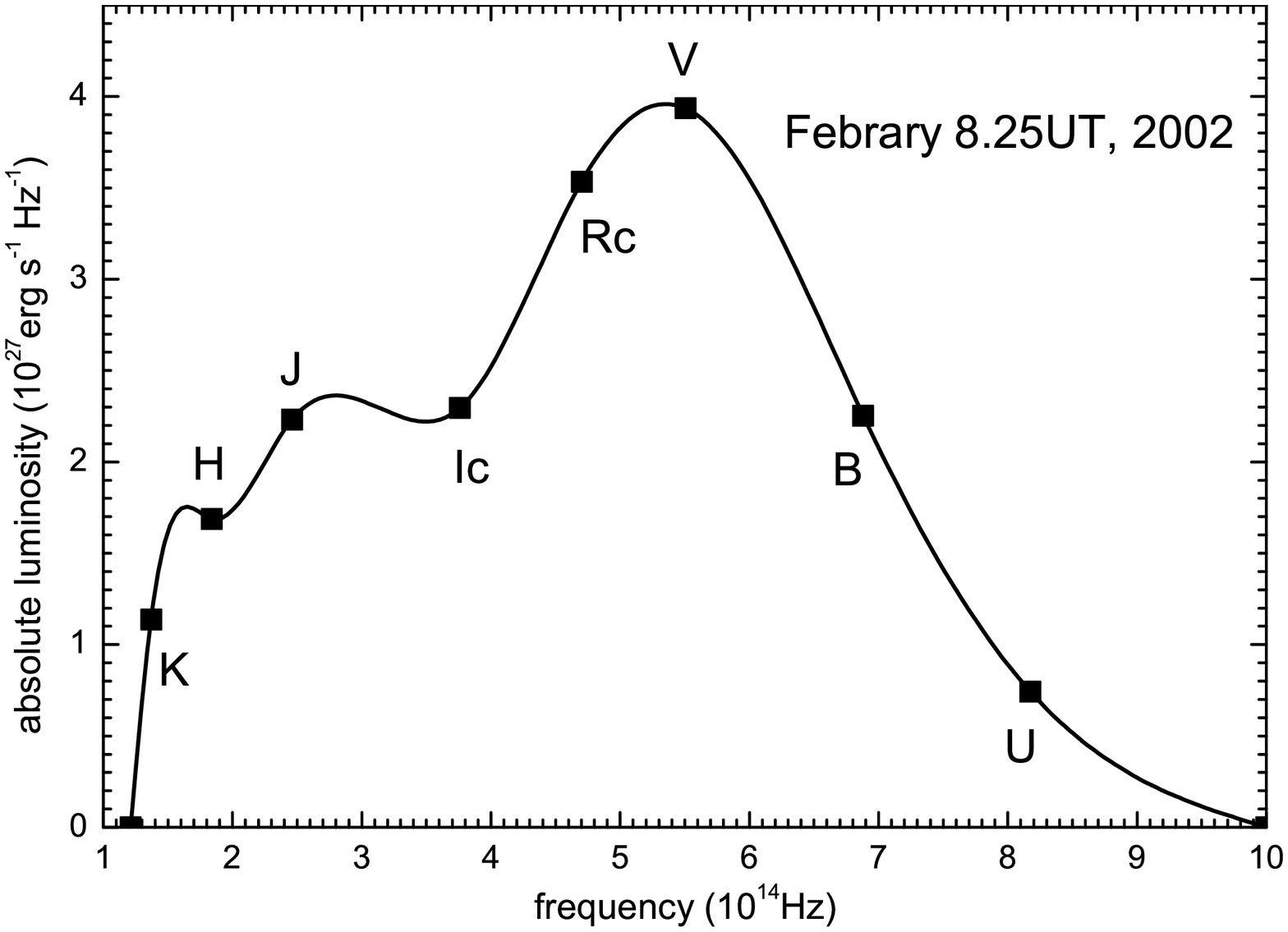]{The monochromatic flux distribution (squares)
of SN~2002ap on February 8, constructed using MAGNUM $UBVRIJHK$
photometry, and the cubic spline fitting curve (solid line).
\label{fig flux} }

\figcaption[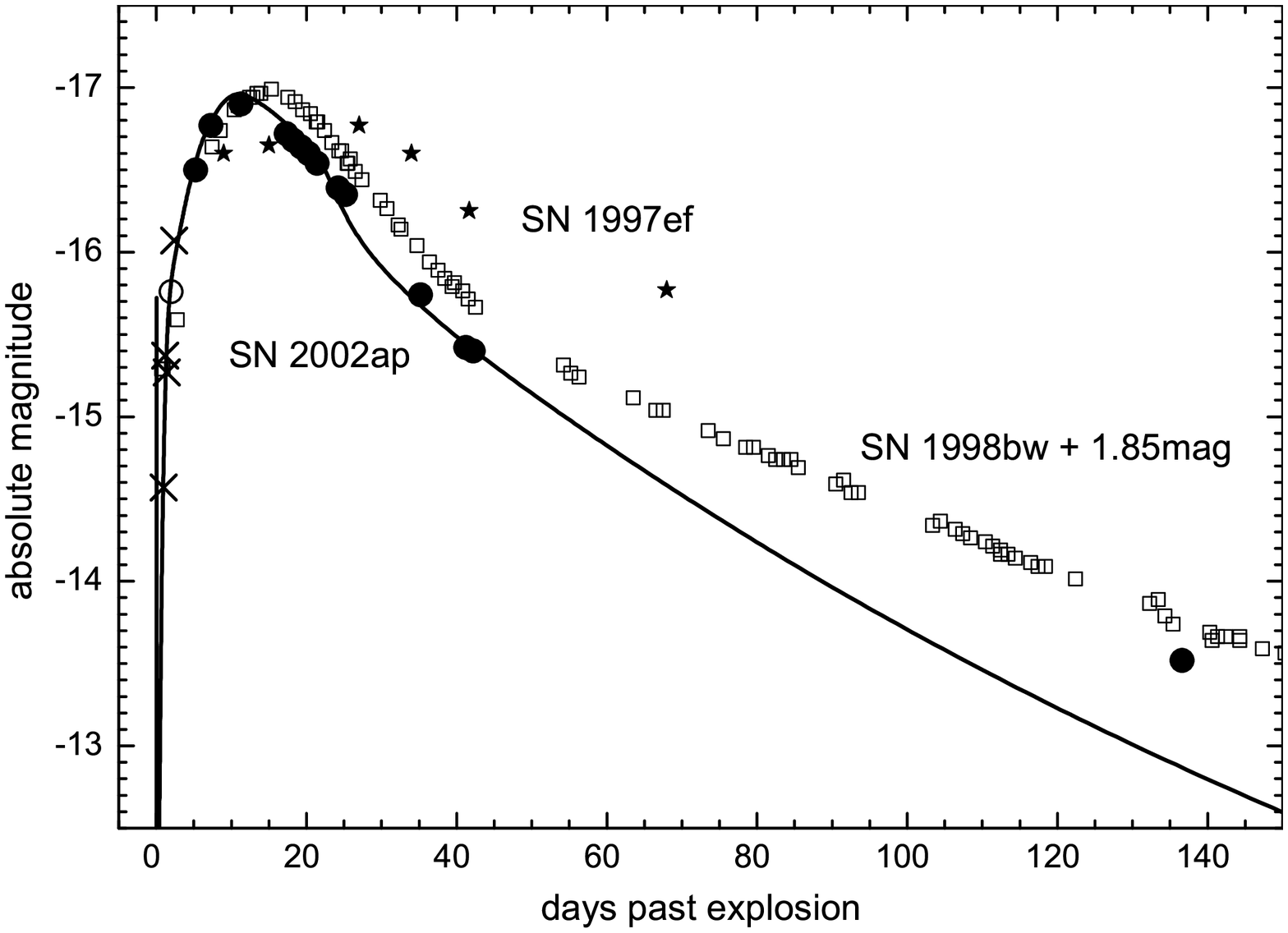]{The bolometric light curve of SN~2002ap,
constructed using MAGNUM $UBVRIJHK$ photometry (filled circles),
compared with the synthetic light curve obtained from a model of a
spherically symmetric explosion with $M_{\rm ej}=2.5~M_\sun$ and
$E_{\rm K}=4\times 10^{51}~{\rm ergs}$ (solid line). Also shown
are one spectrophotometric point based on a Monte Carlo synthetic
spectrum for a date earlier than MAGNUM's first data point (open
circle), and four early unfiltered CCD magnitude points (crosses).
Open squares show the bolometric light curve of SN~1998bw (Patat
et al. 2001) shifted down by 1.85 mag, and filled stars are the
bolometric light curve of SN~1997ef (Mazzali, Iwamoto \& Nomoto
2000). \label{fig lc_model} }

\figcaption[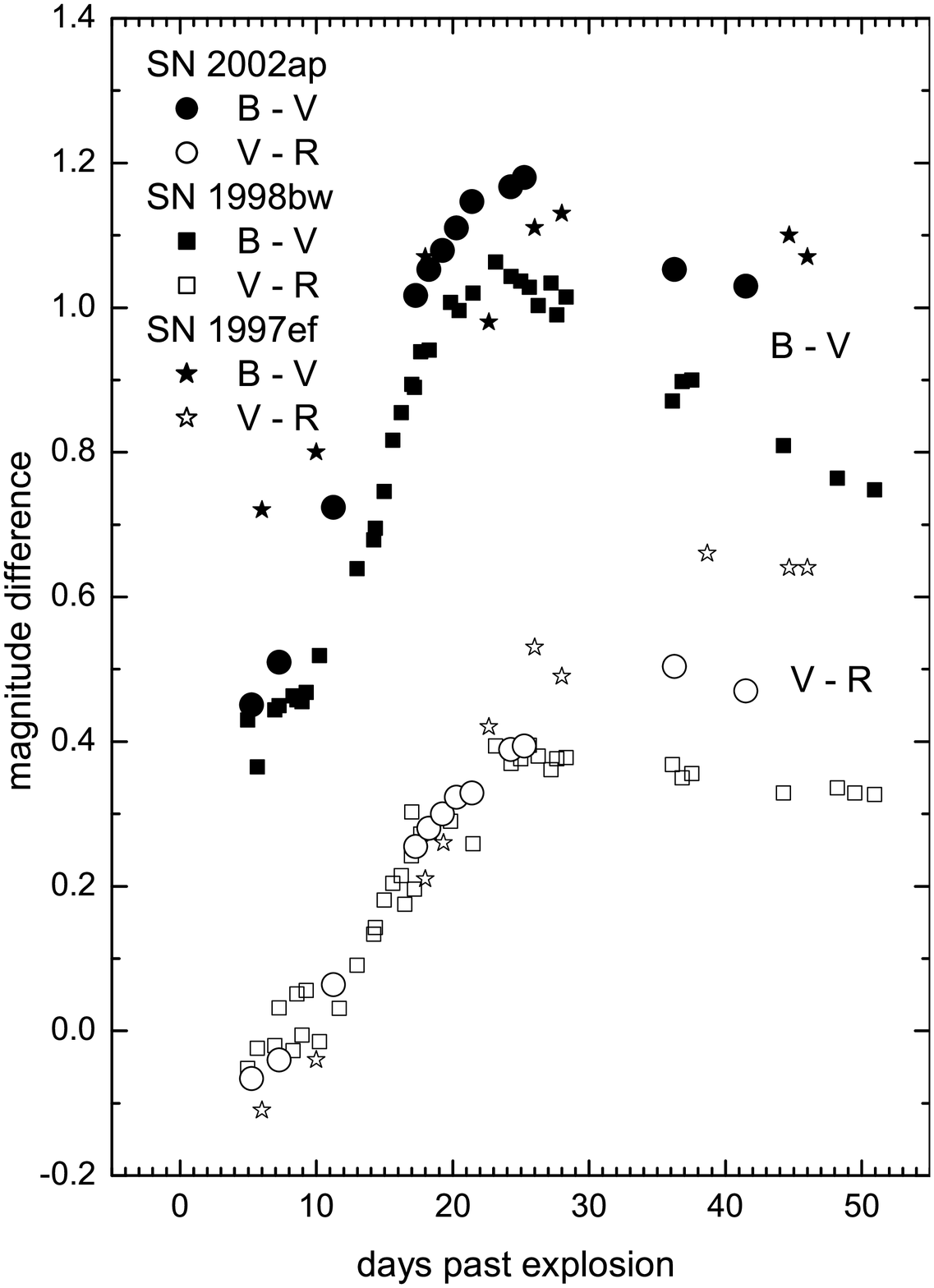]{Evolution of $B-V$ and $V-R$ of the three Type
Ic hypernovae, SNe~2002ap, 1998bw and 1997ef. The time axes
of SNe~1998bw and 1997ef are shorted by a factor of 1.5.
The data of SN~1998bw are calculated from the photometry of
Patat et al. (2001) and those of SN~1997ef are evaluated from
the spectra of Mazzali et~al. (2000).
\label{fig color} }

\figcaption[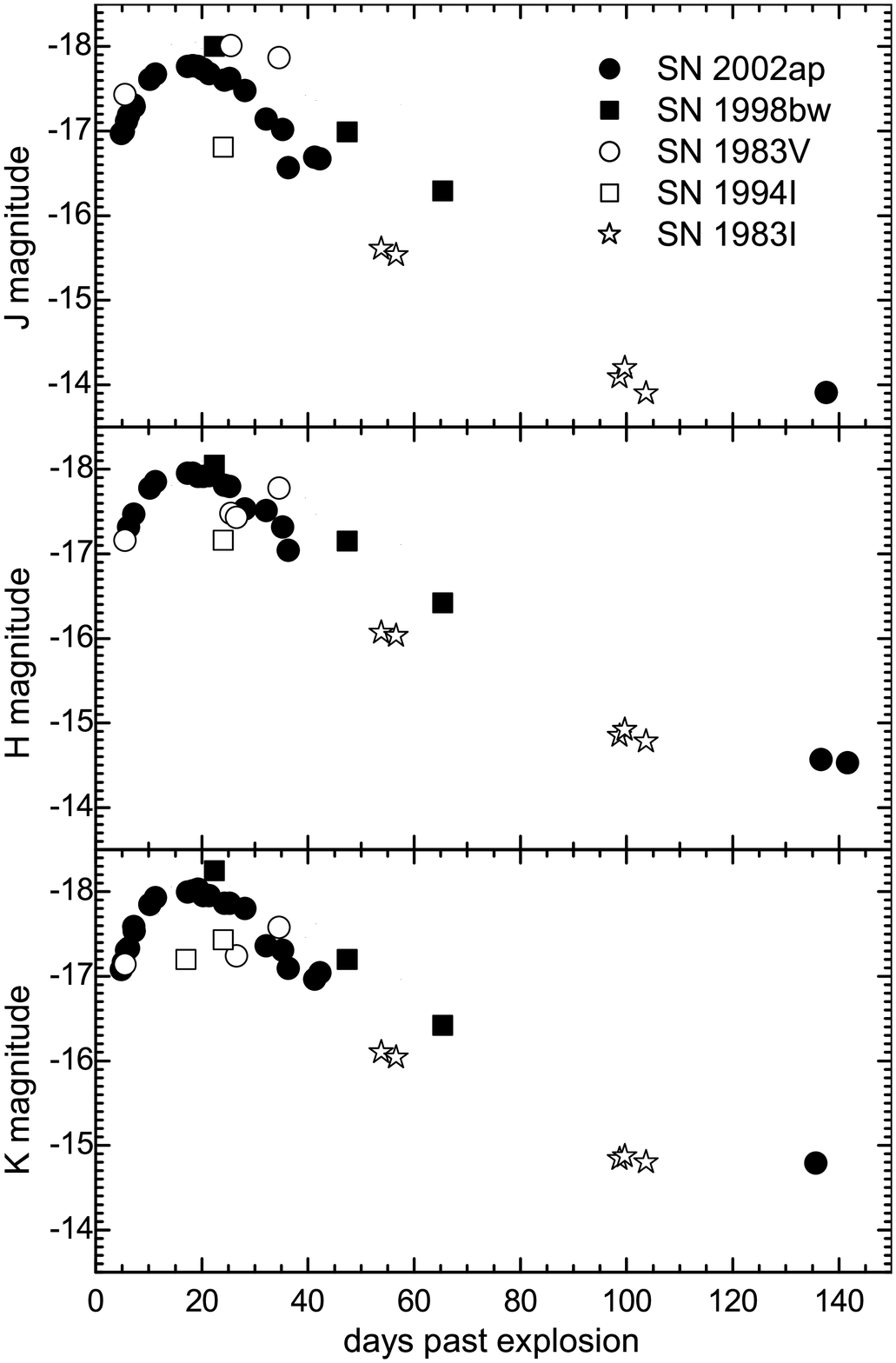]{Comparison of the $J$ (top), $H$ (middle)
and $K$ (bottom) light curves in absolute magnitudes of the 5
Type Ic SNe 2002ap, 1998bw, 1983V, 1994I and 1983I.
\label{fig NIR} }



\clearpage

\begin{figure}
\epsscale{.75}
\plotone{f1.eps}
\end{figure}

\clearpage

\begin{figure}
\epsscale{0.9}
\plotone{f2.eps}
\end{figure}

\begin{figure}
\epsscale{1}
\plotone{f3.eps}
\end{figure}

\clearpage

\begin{figure}
  \plotone{f4.eps}
\end{figure}

\clearpage

\begin{figure}
  \plotone{f5.eps}
\end{figure}

\clearpage

\begin{figure}
\epsscale{.90}
\plotone{f6.eps}
\end{figure}


\clearpage

\begin{deluxetable}{llccccccccc}
\tabletypesize{\scriptsize}
\rotate
\tablecaption{Data of light curves of
SN 2002ap} \tablewidth{0pt} \tablenum{1} \tablehead{ \colhead{UT}
& \colhead{MJD} & \colhead{$m_U$}  & \colhead{$m_B$} &
\colhead{$m_V$}  & \colhead{$m_R$}  & \colhead{$m_I$}  &
\colhead{$m_J$}  & \colhead{$m_H$}  & \colhead{$m_K$}  & Weather }
\startdata
  Feb 2.25 & 52307.25 &  13.39$\pm 0.05$ & 13.36$\pm 0.02$ & 12.82$\pm 0.01$ &
           12.80$\pm 0.02$ &        ...         &  12.59$\pm 0.04$ & ...
                           & 12.36$\pm 0.06$ & clear \nl
  Feb 4.25 & 52309.25 &  13.19$\pm 0.05$ & 13.15$\pm 0.02$ & 12.55$\pm 0.01$ &
           12.51$\pm 0.02$ &       ...          &  12.29$\pm 0.04$ &
                ...           & 11.99$\pm 0.05$ & clear \nl
  Feb 8.25 & 52313.25 &  13.58$\pm 0.05$ & 13.20$\pm 0.02$ & 12.38$\pm 0.01$ &
           12.24$\pm 0.02$ &      ...           &  11.90$\pm 0.02$ &
           11.70$\pm 0.03$ & 11.60$\pm 0.03$ & clear \nl
  Feb 14.27 & 52319.27 &     ...            & 13.76$\pm 0.02$ & 12.65$\pm 0.01$ &
           12.32$\pm 0.02$ & 12.32$\pm 0.01$ &  11.81$\pm 0.03$ &
           11.60$\pm 0.02$ & 11.53$\pm 0.03$ & clear \nl
  Feb 15.25 & 52320.25 &  14.63$\pm 0.06$ & 13.86$\pm 0.02$ & 12.72$\pm
0.01$ &
           12.36$\pm 0.02$ & 12.34$\pm 0.01$ &  11.81$\pm 0.02$ &
           11.60$\pm 0.03$ & 11.52$\pm 0.02$ & clear \nl
  Feb 16.25 & 52321.25 &  14.76$\pm 0.06$ & 13.95$\pm 0.02$ & 12.78$\pm
0.01$ &
           12.40$\pm 0.02$ & 12.35$\pm 0.01$ &  11.82$\pm 0.02$ &
           11.63$\pm 0.02$ & 11.50$\pm 0.03$ & clear \nl
  Feb 17.25 & 52322.25 &  14.84$\pm 0.07$ & 14.05$\pm 0.02$ & 12.85$\pm
0.01$ &
           12.45$\pm 0.02$ & 12.38$\pm 0.01$ &  11.85$\pm 0.02$ &
           11.63$\pm 0.02$ & 11.58$\pm 0.04$ & clear \nl
  Feb 18.4 & 52323.40 &   14.92$\pm 0.07$ & 14.15$\pm 0.02$ & 12.91$\pm
0.01$ &
           12.50$\pm 0.02$ & 12.41$\pm 0.01$ &  11.90$\pm 0.02$ &
           11.63$\pm 0.02$ & 11.58$\pm 0.02$ & clear \nl
  Feb 21.24 & 52326.24 &  15.18$\pm 0.05$ & 14.40$\pm 0.02$ & 13.14$\pm
0.01$ &
           12.67$\pm 0.02$ & 12.51$\pm 0.01$ &  11.98$\pm 0.03$ &
           11.74$\pm 0.03$ & 11.67$\pm 0.05$ & clear \nl
  Feb 22.24 & 52327.24 &  15.28$\pm 0.05$ & 14.48$\pm 0.02$ & 13.21$\pm
0.01$ &
           12.74$\pm 0.02$ & 12.55$\pm 0.01$ &  11.96$\pm 0.02$ &
           11.76$\pm 0.04$ & 11.67$\pm 0.03$ & clear \nl
  Mar 4.23 & 52337.23 &   15.95$\pm 0.08$ &    ...            & 14.00$\pm
0.01$ &
                  ...         & 12.98$\pm 0.01$ &  12.56$\pm 0.06$ &
           12.23$\pm 0.05$ & 12.24$\pm 0.37$ & thin cloud \nl
  Mar 5.24 & 52338.24 &   16.08$\pm 0.09$ & 15.23$\pm 0.02$ & 14.08$\pm
0.01$ &
           13.50$\pm 0.02$ & 13.04$\pm 0.01$ &  ... &
           ... & ... & thin cloud \nl
  Mar 10.24 & 52343.24 &       ...          &     ...            &  14.36$\pm
0.01$ &
           13.80$\pm 0.02$ & 13.24$\pm 0.01$ &  12.89$\pm 0.06$ &
                ...           & 12.57$\pm 0.08$ & thin cloud \nl
  Mar 11.23 & 52344.23 &     ...            & 15.48$\pm 0.04$ & ...
  &
           13.83$\pm 0.02$ & 13.27$\pm 0.01$ &  12.91$\pm 0.06$ &
                ...           & 12.53$\pm 0.12$ & thin cloud \nl
  Jun 12.61 & 52437.61 &     ...            & ... & 16.14$\pm 0.03$
  &  ...  & ...  &  ...  &  ...   & 15.01$\pm 0.07$ & clear \nl
  Jun 13.60 & 52438.60 &     ...            & 16.78$\pm 0.04$ & ...
  &  ...  & ...  &  ...  &  14.90$\pm0.06$   &  ...  & clear \nl
  Jun 14.60 & 52439.60 &     ...            &  ...  & ...
  & 15.41$\pm 0.03$ &  ...  &  15.72$\pm 0.06$ &  ...
  &  ...  & clear \nl
  Jun 16.60 & 52441.60 &     ...            &  ...  & ...
  & ...  & 15.08$\pm 0.04$ &  ...  &  ...  &  ...  & clear \nl
  Jun 18.60 & 52443.60 &     ...            &  ...  & ...
  &  ...  & 15.12$\pm 0.03$ &  ...  &  15.05$\pm0.06$
  &  ...  & clear \nl

\enddata
\tablecomments{Apparent magnitudes were measured within the
aperture size of 11 arcsec, but a smaller aperture was adopted to
enhance the S/N ratio for the $U$-band images taken on March $4-5$
(8.3 arcsec) and the images taken in June (5.5 arcsec for $BVRI$,
and 8.3 arcsec for $JHK$)  }
\end{deluxetable}

\clearpage

\begin{deluxetable}{lcccccccc}
\tabletypesize{\scriptsize}
\tablecaption{Characteristics of light curves of SN 2002ap}
\tablewidth{0pt}
\tablenum{2}
\tablehead{
\colhead{Item} & \colhead{$U$ band}  &
\colhead{$B$ band}  & \colhead{$V$ band}  &
\colhead{$R$ band}  & \colhead{$I$ band}  &
\colhead{$J$ band}  & \colhead{$H$ band}  & \colhead{$K$ band}
}
\startdata
  Peak date (UT) &  Feb 4.6$\pm 2$ & Feb 6.0$\pm 1$ &
         Feb 7.9$\pm 2$ &  Feb 9.9$\pm 1$ & Feb 12.7$\pm 2$ &
         Feb 12.9$\pm 2$ &  Feb 13.0$\pm 2$ & Feb 12.9$\pm 2$ \nl
  Peak date (MJD) &  52309.6$\pm 2$ & 52311.0$\pm 1$ &
         52312.9$\pm 2$ &  52314.9$\pm 1$ & 52317.7$\pm 2$ &
         52317.9$\pm 2$ &  52318.0$\pm 2$ & 52317.9$\pm 2$ \nl
  Peak app mag  &  13.20 & 13.11 & 12.38 & 12.22 &
         12.31 & 11.79  &  11.59 & 11.51 \nl
  Peak abs mag  & -16.30 & -16.39 & -17.12 & -17.28 &
         -17.19 & -17.71  &  -17.91 & -17.99 \nl
  Gradient (mag/day) &  0.08$\pm 0.01$ &
         0.08$\pm 0.01$ & 0.08$\pm 0.01$ & 0.07$\pm 0.01$ &
         0.04$\pm 0.01$ & 0.06$\pm 0.01$ & 0.05$\pm 0.01$ & 0.06$\pm 0.01$ \nl
\enddata
\tablecomments{Peak date and apparent magnitude in each band are
   determined from the cubic spline fitting to the data.  Peak
   absolute magnitude is based on the distance modulus $m-M=29.5$
   mag (Sharina et al. 1996; Sohn \& Dvidge 1996).  The decline rate
   in each band was determined from linear fitting to the data taken well
   after the peak date
  (Feb 16 - Mar 5 for $U$ and $V$,
   Feb 15 - Mar 5 for $B$,
   Feb 21 - Mar 5 for $R$ and $I$,
   Feb 22 - Mar 5 for $J$, $H$ and $K$).
   }
\end{deluxetable}

\begin{deluxetable}{lccc}
\tabletypesize{\scriptsize}
\tablecaption{Bolometric magnitudes of
SN~2002ap
              and bolometric corrections}
\tablewidth{0pt}
\tablenum{3}
\tablehead{
\colhead{UT} & \colhead{MJD}  &
\colhead{$M_{\rm bol}$} & \colhead{B.C.}
}
\startdata
  Feb 2.25  & 52307.25 & -16.50 & 0.45 \nl
  Feb 4.25  & 52309.25 & -16.77 & 0.45 \nl
  Feb 8.25  & 52313.25 & -16.90 & 0.49 \nl
  Feb 14.27 & 52319.27 & -16.72 & 0.40 \nl
  Feb 15.25 & 52320.25 & -16.68 & 0.37 \nl
  Feb 16.25 & 52321.25 & -16.64 & 0.35 \nl
  Feb 17.25 & 52322.25 & -16.60 & 0.32 \nl
  Feb 18.40 & 52323.40 & -16.54 & 0.32 \nl
  Feb 21.24 & 52326.24 & -16.39 & 0.24 \nl
  Feb 22.24 & 52327.24 & -16.35 & 0.21 \nl
  Mar 4.23  & 52337.23 & -15.74 & 0.03 \nl
  Mar 10.24 & 52343.24 & -15.42 & -0.01 \nl
  Mar 11.23 & 52344.23 & -15.40 & 0.01 \nl
  Jun 13.6  & 52438.6  & -13.52 & 0.11 \nl
\enddata
\end{deluxetable}

\end{document}